\documentclass[aps,noshowkeys,noshowpacs,superscriptaddress,twocolumn,floatfix]{revtex4}

\usepackage{psfrag}
\usepackage{graphicx}
\usepackage{amsmath}
\usepackage{hyperref}
\usepackage{color}

\begin{document}
 
\title{Communication and trust in the bounded confidence model}
 
\author{M.~J.~Krawczyk}
\affiliation{
Faculty of Physics and Applied Computer Science, AGH University of Science and Technology, al. Mickiewicza 30, PL-30059 Krak\'ow, Poland
}

\author{K.~Malarz}
\homepage{http://home.agh.edu.pl/malarz/}
\email{malarz@agh.edu.pl}
\affiliation{
Faculty of Physics and Applied Computer Science, AGH University of Science and Technology, al. Mickiewicza 30, PL-30059 Krak\'ow, Poland
}

\author{R.~Korff}
\affiliation{
Mainland Southeast Asian Studies, Universit\"at Passau, Innstrasse 43, 94032 Passau, Germany
}

\author{K.~Ku{\l}akowski}
\affiliation{
Faculty of Physics and Applied Computer Science, AGH University of Science and Technology, al. Mickiewicza 30, PL-30059 Krak\'ow, Poland
}

\date{\today}

\begin{abstract}
The communication process in a situation of emergency is discussed within the Scheff theory of shame and pride. The communication involves messages from media and from other persons. Three strategies are considered: selfish (to contact friends), collective (to join other people) and passive (to do nothing). We show that the pure selfish strategy cannot be evolutionarily stable. The main result is that the community structure is statistically meaningful only if the interpersonal communication is weak.
\end{abstract}

\maketitle
 
\section{Introduction}

Communication is a key process in social systems and it is of interest from various perspectives. For a political scientist the feedback between media and the audience in democratic systems determines the outcome of political decisions \cite{1}. For a system-oriented sociologist, communication enables a social system to maintain its identity by reduction the complexity of its environment \cite{2}. In between, in social psychology, people communicate to create meanings and these meanings determine human actions \cite{3}.  In psychology, communication is investigated in terms of transactional analysis \cite{4}, a Freudist scheme to decipher human attitudes. In mathematics, the physical concept of entropy was generalized within communication theory \cite{5}, which is at the foundations of computer science. In game theory, communication allows for new solutions in non-zero-sum games which yet remain unsolved \cite{6}. It makes sense to look for interdisciplinary models of the communication processes, which could integrate at least some of mathematical formulations used in different fields. 

In theory of Artificial Intelligence, Hebbian learning theory \cite{7} and the neurological perspective \cite{8} were used to equip simulated agents with human-like beliefs and emotions \cite{9}. The outcome of this formulation was a set of differential equations, where the time-dependent variables were the level of belief,  the stimulus, the feeling and the preparation of the body state. The relations between these variables were assumed to depend on some additional parameters, as learning rate from feeling to belief etc. A detailed description of these models can be found in \cite{9,9a}. A similar model of dynamics of public opinion was formulated by Zaller \cite{1} in terms of master equations for conditional probabilities. There, a set of independent agents was subject to a stream of messages coming from media. The variables used were the probabilities that messages are received and that they are accepted; additional parameters were introduced to describe the credibility of messages, the awareness on resistance to persuasion, the predisposition on resistance to persuasion etc. A concise description of this mathematical formulation can be found in Section II of \cite{10}. The Zaller approach to theory of public opinion was reformulated in \cite{10} to a geometrical scheme, similar to the bounded confidence model \cite{11p}. In this new scheme, an agent was represented by a set of messages he received in the past. The messages were expressed as points on a plane of issues, and the probability that a message was received by an agent was postulated to depend on the position of the point with respect to the messages received by him previously. Later, the communication between agents was added to the model \cite{11}. 

The aim of this paper is to use the same geometrical model \cite{10,11} to describe how trust emerges from beliefs of agents and the communication between them \cite{9}. A new element here is the time dependence of the threshold value, which measures the trust between agents. We adopt the definition of trust formulated in \cite{12}: ``{\em trust is an emotional attitude of an agent towards an information source (e.g. another human agent, an ambient device) that determines the extent to which information received by the agent from the source influences agent's belief(s)}''.  This formulation is close to the idea of conditional probability used by Zaller \cite{1}, that the information is accepted. Similar time-dependent couplings between the probabilities of different actions of agents were found to be efficient in modeling  social systems \cite{13}. Also, the influence of one agent on the action of another one is a direct reference of the sociological definition of power \cite{14}; here we interpret it as an enhancement of a positive self-evaluation or the receiver by sending to him a message which he is willing to accept.

The sociological context we bear in mind is that some amount of agents is faced with an unexpected, dangerous situation. The degree of the danger can vary from a direct threat, as an evacuation of people from a building or a train after an explosion, to an anxiety of ship passengers of an unverified possibility of storm. In any case, agents can remain passive or they can act in a selfish way or they can join in a collective action. Each solution brings some risk and inconvenience, and the necessity to decide is itself an unwanted circumstance, which leads to stress and lowers the agents' self-estimation. We intend to discuss their dilemma in terms of the Scheff theory of pride and shame \cite{15}, which seems to provide a set of ideas which are particularly appropriate for our emergency scenario.

This text is organised as follows. In the next section we provide a short description of the emergency scenario in terms of the Scheff theory. 
Section \ref{Sec-3} is devoted to a mathematical formulation of two models, which are used here to evaluate the efficiency of contacting with relatives or friends by phone; doing this is considered here to be the selfish strategy. In the same section we demonstrate numerical results, which state the conditions when this selfish strategy is not helpful. Section \ref{Sec-4} defines the Zaller--Deffuant model of bounded confidence \cite{11p,11} with the time-dependent trust levels. The main result of this paper is the time dependence of the structure of communities of the evacuated people, described in Section \ref{Sec-5}. This dynamics is governed by the communication process. Last section is devoted to conclusions.

\begin{figure}[ht]
\psfrag{c5}{$p_2$}
\psfrag{lambda, pprim}{$\lambda, p'$}
\psfrag{p}{$p$}
\includegraphics[width=0.49\textwidth]{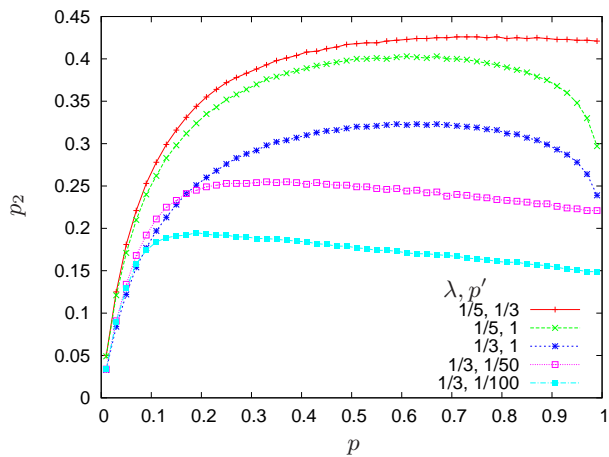}
\caption{The probability of a successful talk $p_2(p)$ against the probability $p$ of intention to talk, obtained within the cellular automaton approach. The size of the lattice is $L\times L$, where $L=10^3$. The parameters $p'$ and $\lambda$ are the probabilities of stopping unsuccessful and successful connection, respectively.}
\label{fig-1}
\end{figure}

\begin{figure}[ht]
\psfrag{p2}{$p_2$}
\psfrag{p}{$p$}
\psfrag{ks, tau}{$k_s$, $\tau$}
\includegraphics[width=0.49\textwidth]{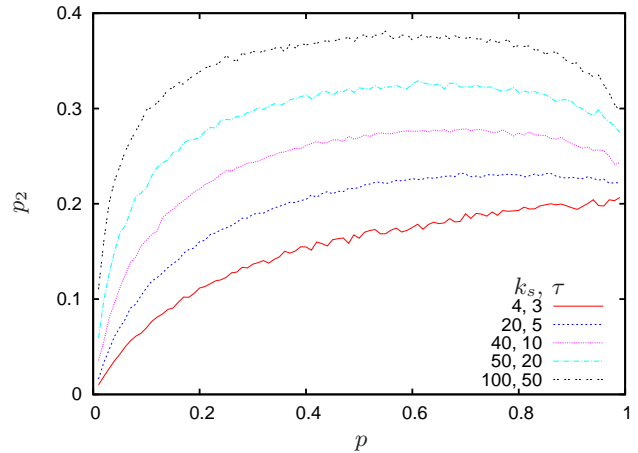}
\caption{The probability of a successful talk $p_2(p)$ against the probability $p$ of intention to talk, obtained within the network approach. The size of the network is $N=100$ except the lowest curve, where $N=1000$. The results are the same for larger $N$, except that the fluctuations decrease. The parameter $k_s$ is the mean degree of the Erd{\H o}s--R\'enyi network, and the parameter $\tau$ is the mean time of successful talks.}
\label{fig-2}
\end{figure}

\begin{figure}[ht]
\psfrag{mu=}{$\mu=$}
\psfrag{t}{$t$}
\psfrag{average(a(i,j))}{$\langle a_{i,j} \rangle$} 
\includegraphics[width=0.49\textwidth]{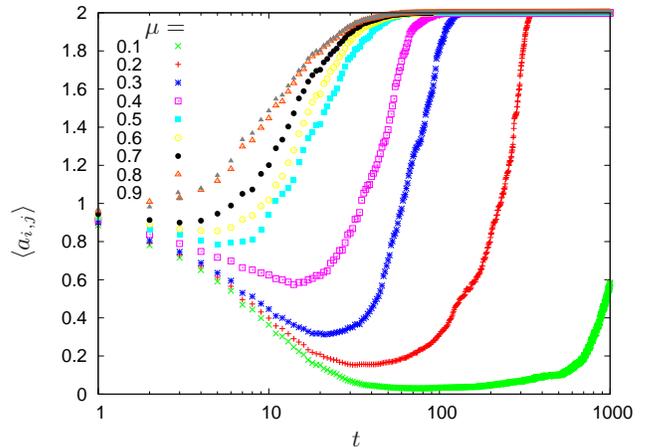}
\caption{The time dependence of the mean value $\langle a_{i,j}\rangle$ of the trust matrix element normalized to the range $2\mu$ for different values of the parameter $\mu$. }
\label{fig-3}
\end{figure}

\begin{figure}[ht]
\psfrag{Q}{$Q$}
\psfrag{t}{$t$}
\includegraphics[width=0.49\textwidth]{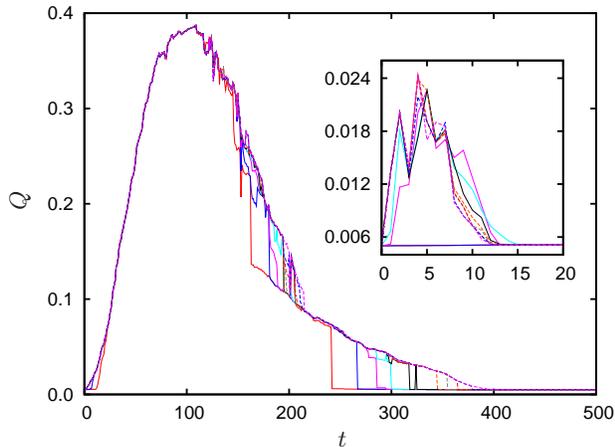}
\caption{The time dependence of the modularity $Q$ for $\mu=0.2$ and different values of the parameter $\beta$ according to the method of differential equations \cite{21}. In the inset the same plot for $\mu=0.7$.}
\label{fig-4}
\end{figure}

\begin{figure}[ht]
\psfrag{Qmax}{$Q_{\text{max}}$}
\psfrag{mu}{$\mu$}
\psfrag{Newman}{Ref. [21]}
\psfrag{RR}{Ref. [22]}
\includegraphics[width=0.49\textwidth]{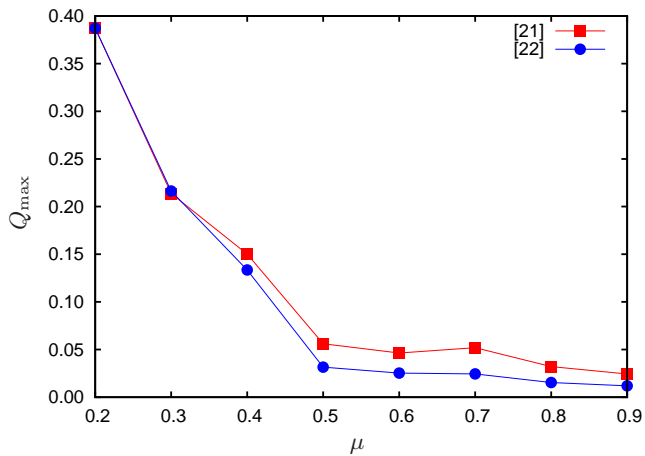}
\caption{The maximal value of the time dependence of the modularity $Q(t)$ against the tolerance parameter $\mu$ according to the Newman algorithm \cite{19} (red curve) and the algorithm of differential equations \cite{20} (blue curve). }
\label{fig-5}
\end{figure}

\section{\label{Sec-2}Scheff theory of an emergency scenario}

A catastrophe, a terrorist attack, or any other strong threat is obviously a strong disturbance of current plans of its participants,
accompanied perhaps with a direct threat for their health or even life. As such, it must give raise to strong negative emotions, triggered by fear. Here we are interested in the communication between the participants. For each one, there are three options: {\it i}) to communicate by phone with friends or relatives, who do not participate in the situation, {\it ii}) to contact with other participants and in a search for a collective strategy, or {\it iii}) to withdraw and do nothing. These three strategies will be termed as S (selfish), C (collective) and P (passive), respectively. It might be surprising that during the 7/7 London bombings, the passive strategy was observed as a quite common \cite{16}.

According to the theory of pride and shame by Thomas J. Scheff \cite{15}, the strategy selected by the participants depends largely on conscious or unconscious messages which they send and receive. Namely, they try to evaluate how they are perceived by other participants: with respect or not. In the former case their positive self-estimation is enhanced, what enables the interpersonal attunement and the social solidarity. Consequently, they are willing to select the strategy C, what stabilizes the mutual respect. This cycle can be termed as the loop of pride. On the contrary, disrespectful signals lead to an enhancement of the negative self-evaluation, a shame and its repression and a hostility. As a consequence the lack of respect is stabilized and chances for C decrease; what is being selected is S or P. This behaviour can be termed as the loop of shame. The scheme of two loops in the Scheff theory was described in \cite{17}.

As we see, a positive feedback is present in both loops. However a switching is possible from the stage of shame in the loop of shame to the stage of an interpersonal attunement in the loop of pride \cite{17}. This switching is possible by means of a realization of the shame. According to Scheff, it is an unconscious shame what breaks interpersonal bonds. Consequently, getting in contact with other participants of the scenario helps to realize, that the threat is common and the solidarity can be restored. Below we investigate the dynamics of the possible spread of this realization, from those who are already conscious to those who are not yet. Once they get into the loop of pride, they propagate this state further.

\section{\label{Sec-3}Models of a phone communication jam}

The selfish strategy S is natural in the case of large catastrophes, where everybody wants to get know if the relatives or friends remain alive. However, the payoff heavily depends on the amount of persons who select this strategy. It is straightforward to expect that once the amount of phone speakers excesses some percentage of the population, successive attempts to contact by phone must fail even if the phone network works properly. It is just almost impossible to have simultaneously more than one phone call. We demonstrate the effect by an evaluation of the percentage $p_2$ of agents who successfully talk to each other, and not only try to get connection; the probability of the latter state is denoted as $p_1$ from now on. These probabilities are to be calculated against the percentage $p$ of people who try to get connection.

The calculation is performed within two different approaches. First one is a cellular automaton with parallel updating. A single variable $s_{i,j}$ is assigned to each site of the square lattice with the helical boundary conditions. The list of possible states of $s_{i,j}$ consists $0, L, R, U, D, L^*, R^*, U^*$ and $D^*$. Agents in the state `0' are silent. With the probability $p$ an agent picks up his phone and starts to calling to his/her randomly selected neighbour. Agents in states $L,R,D,U$ are trying to connect to his/her nearest neighbour situated on left, right, down or up, respectively. They stop unsuccessful attempts to connect with probability $p'$. Agents in states $L^*,R^*,D^*,U^*$ talk with his/her left, right, down or upper neighbour, respectively. They finish their talk and return to the state `0' with the probability $\lambda=1/\tau$. The probability $p_1$ is calculated as the amount of agents in one of the states $L,R,D,U$, while the probability $p_2$ is calculated as the amount of agents in one of the states $L^*,R^*,D^*,U^*$.

In the second approach, agents are distributed at nodes of Erd{\H o}s--R\'enyi network, and the updating is sequential. The variables used $w_{i,j}$ deal with the states of bonds between agents; $w_{i,j}=1$ means that $i$ tries to phone to $j$, while $w_{i,j}=0$ means that he does not try. The probability $p_2$ is found as the percentage of agents involved in mutual talks. The probability $p_1$ is the amount of agents $i$ such that $w_{i,j}=1$, and $w_{j,i}=0$. The simulation goes as follows. Once $i$ is selected randomly with repetitions, we ask if there is a neighbour $j(i)$ such that $w_{i,j}=1$. If yes, we check if $w_{j,i}=1$. If yes, $i$ and $j$ are talking; the talk is broken with probability $\lambda$. Coming back, if $w_{j,i}=0$, we check if there is any neighbour $k(j)$ such that $w_{j,k}=1$. If not, $w_{j,i}$ is set to one; this means that $j$ starts to talk with $i$. If yes, $w_{i,j}$ is set to zero; then $i$ selects randomly one of his neighbours $n$ and tries to connect him with probability $p$; again $w_{i,n}$ is set to one. Next possibility about $i$ is that $w_{i,j}=0$ for all his neighbours $j(i)$. Then, if any $w_{j,i}=1$, $w_{i,j}$ is set to one. If $w_{j,i}=0$ for all $i$'s neighbours $j(i)$, again $i$ tries to communicate with one of his neighbours.

These approaches are very different in details. In the cellular automaton we use von Neumann neighborhoods of four cells, and the topology is just a plane. On the contrary, in the case of the Erd{\H o}s--R\'enyi network the topology is random with the small world property. Also, in the case of network unsuccessful calls initiated in one step are stopped only after second selection of the same node. Still, as we see in Figs. 1 and 2, the result is qualitatively the same: in both models the curves $p_2(p)$ do not increase above some value $p^*$ of $p$.  As this result is obtained within two entirely different models, it can be considered as validated. Although we do not specify the payoffs, it is clear that the strategy S can be efficient only if it is chosen by a minority. This effect allows to expect that other strategies, P or C, can be active.

\section{\label{Sec-4}Bounded confidence model}

The model of communication used here is a slightly modified version of the Zaller--Deffuant model \cite{11}.  There are two kinds of messages: those from media and those from the agents themselves. All messages are represented by points on a plane. A new message is accepted by an agent if the position of the message is not too far from the messages accepted by this agent in the past. This means that the system is characterized by a critical distance $\mu$ in the message space; `not too far' means `the distance is shorter, than $\mu$'. In the original formulation of the bounded confidence model \cite{11p}, the parameter $\mu$ meant the tolerance for distant opinions. We are willing to maintain this meaning of $\mu$ in this formulation.

The system is subject to a stream of messages from media. Besides of that, each agent reproduces
one of messages accepted by him previously; he does so with the probability $r$, which we select to be $r=10/N$, where $N$ is the number of agents \cite{11}. The modifications of the model with respect to the original version \cite{11} are, that {\it i}) we do not distinguish between the messages received and accepted---this can be done, because we do not discuss final decisions of the agents;  {\it ii}) agents send particular messages, and not their time averages; {\it iii}) what is most important, the value of the critical distance $a_{i,j}$ depends both on the sender $j$ and on the receiver $i$ and it varies in time; in this way we formalize the definition of trust, given in \cite{12}. On the contrary to this variation, the critical distance $\mu$ for the messages sent by media remains constant.

The time dependence of $a_{i,j}$ is determined by the following rules. The initial values of $a_{i,j}$ are equal to $\mu$. Once a message from $j$ is accepted by $i$, $a_{i,j}$ is transformed to $a_{i,j}/2+\mu$. Once a message is not accepted, $a_{i,j}$ is transformed into $a_{i,j}/2$.  In this way, the matrix elements of trust between agents vary between zero and $2\mu$. In this variation, more recent messages matter more; after three subsequently accepted messages, the respective matrix element $a_{i,j}$ increases from zero to 0.875 of the maximal possible value. In other words, the memory of the system remains finite. These rules were found to be useful in cooperation modeling \cite{13}.

The message positions are limited to a square on a plane $(x,y)$, where both coordinates vary between $-1$ and $+1$. Actually, the critical distance is defined with respect to the size of this square. After a sufficiently long time, each agent is going to accept each message \cite{10,11}; this is assured by the increase of the area around the accepted messages from media. Here we are not interested in the asymptotic regime, but rather in the transient 
process of filling the square of particular agents by accepted messages. The role of the interpersonal communication in this process is encrypted in the time dependence of the trust matrix $a_{i,j}$. At the asymptotic stage, i.e. for time long enough, $a_{i,j}=2\mu$ for all $i,j$. In the transient time, the role of the communication between agents and the role of the messages from media can depend on the value of the threshold $\mu$. We are interested in the structure of communities of agents where the mutual trust is established. These results are described in the next section.

\section{\label{Sec-5}Communities}

The plot shown in Fig. 3 shows the time dependence of the mean value $\langle a_{i,j}\rangle$ of the trust matrix elements for different values of the trust parameter $\mu$. The diagonal matrix elements are excluded. Although there is no clear difference between neighboring curves, two different regimes can be observed. For $\mu \ge 0.7$, the obtained curve increases almost monotoneously with time. On the contrary, for $\mu$ close to 0.2 and nearby, the curve first decreases, later increases to the value $2\mu$. The decrease mark the overall fall of interpersonal trust. At this stage, the process of an increase of individual areas around accepted messages is due mostly to the messages from media. The difference is then between the communication dominated by the interpersonal messages (large $\mu$) and the one dominated by media (small $\mu$).

The structure of the communities of mutual trust is investigated by means of two algorithms, both designed as to identify communities in networks \cite{18}. Here the network nodes are represented by agents, and the weights of the bonds---by the matrix elements $a_{i,j}$. First algorithm is the agglomerative hierarchical clustering method proposed by Mark Newman \cite{19}; it can be applied to symmetric as well as to non-symmetric matrices.
Second algorithm relies on a numerical solution of a set of nonlinear differential equations \cite{20}, where the unknown variables are the same matrix elements $A_{i,j}=a_{i,j}/(2\mu)$. The equations are
\begin{equation}
\frac{dA_{j,k}}{dt}=\Theta(A_{j,k})\Theta(1-A_{j,k})\sum_i\left(A_{j,i}A_{i,k}-\beta \right),
\end{equation}
where $\Theta(x)$ is the step function and $\beta$ is a model parameter, which measures the threshold, above which the product of the matrix elements starts to increase. This method works on symmetric matrices; to apply it, we have to symmetrize the trust matrix $a_{i,j}$. Both methods make use of the modularity $Q$, defined as \cite{21}
\begin{equation}
Q=\frac{1}{2m}\sum_{i,j}\left(A_{i,j}-\frac{k_ik_j}{2m}\right)\delta_{c_i,c_j},
\end{equation}
where $\delta_{i,j}$ is the symbol of the Kronecker delta, $k_i$ is the weighted degree of node $i$ and $m$ is the total number of edges of the network. The search of the maximal value of the modularity allows to find the optimal structure of communities. Simultaneously, the value of $Q$ allows to evaluate the statistical meaningfulness of the obtained structure. A large value of $Q$ (about 0.3 or more) means that the structure differs
remarkably from a random one.

When $Q$ is small, as for example at the beginning of the simulation, the method of differential equations \cite{20} produces one connected cluster. However, when $Q$ is large, both applied methods give almost the same communities. A brief inspection of the obtained data allows to state that once a community appears, it persists, with some new nodes being attached during the process. In our sample of 100 nodes, new communities of 4-9 nodes are born by a separation from the whole mass, and even if modified, they can be identified. However, the most important result is not a particular structure, but its meaningfulness, measured by $Q$. We observe that {\it i}) the time dependence of the modularity $Q$ displays a strong maximum, {\it ii}) its maximal value is meaningful (in the range $0.2<Q<0.4$) only for small value of $\mu$. These results are shown in Figs. 4 and 5. They mean that the community structure undoubtely appears only if the interpersonal communication is weak.

\section{Conclusions} 

Our calculations are related to the problem of an emergency scenario, when individuals select one of three strategies: a selfish communication with persons not involved into the emergency situation, a passivity, or a collective action. We used two simple models to demonstrate, that the selfish strategy cannot be rewarding, if applied by the majority. Next we concentrated on the community structure, which appears when individuals communicate to join in a collective action. The communication is modeled with using the bounded confidence theory, where the parameter of tolerance measures the 
individual ability to accept messages sent by other persons.

We found a qualitative difference between the communication dynamics for small and large tolerance $\mu$. Loosely speaking, a small value of $\mu$ in our model means that people are willing to accept only these messages which are directly close to their own opinions. As shown in Fig. 3, in these conditions the mutual trust falls quickly, and it is only selected messages from media which are accepted. Once the area around a small number of accepted messages starts to widen, some neighbours can be found in a direct neighbourhood. The reconstructed trust refers only to a few neighbours, and it is strengthened
often by a repetition of mutually copied messages. On the contrary, for large tolerance $\mu$ many interpersonal messages are accepted immediately, and the initial weak fall of the mutual trust is followed by its fast increase, as shown in Fig. 3. In these conditions the connections between agents include the large majority to the same cluster; as they are equally strong, communities are practically absent.

Drawing more from sociology, these conclusions can be supplemented by a note on theory of attributions \cite{17}. People are likely to infer on the causes of events which happen to them. There are two kinds of attributions: internal, when we look for causes in our individual personalities, dispositions and attitudes, and external, when we identify causes in an external world. More than often we apply the former to our successes and the latter---to our defeats \cite{22}. In a society divided into small communities without a communication between them, these negative external attributions are strengthened even more. In this case the collective action of groups can be mutually hostile \cite{23}.

\acknowledgements{One of the authors (K.K.) is grateful to Alexei Sharpanskykh for useful discussions. The research is partially supported within the FP7 project SOCIONICAL, No. 231288.}


\begin{thebibliography}{99}

\bibitem{1} J. R. Zaller, The Nature and Origins of Mass Opinion, Cambridge UP, Cambridge 1992.
\bibitem{2} N. Luhmann,  A Sociological Theory of Law, London: Routledge, 1985. 
\bibitem{3} J. M. Charon, Symbolic Interactionism: An Introduction, An Interpretation, An Integration, Pearson, Prentice Hall, New Jersey, 2010. 
\bibitem{4} I. Stewart and V. Joines, Transactional Analysis Today, A New Introduction to Transactional Analysis,  Lifespace Publ., Nottingham, 2000. 
\bibitem{5} C. E. Shannon, A mathematical theory of communication, Bell Syst. Tech. J. {\bf 27} (1948) 379; {\em ibid} 623.
\bibitem{6} P. D. Straffin, Game Theory and Strategy, Mathematical Association of America, Washington DC, 1993.
\bibitem{7} D. O. Hebb, The Organisation of Behavior, Wiley, NY 1949.
\bibitem{8} A. Damasio, Descartes' Error: Emotion, Reason, and the Human Brain, Putnam, NY 1994.
\bibitem{9} Z. A. Memon and J. Treur, Modeling the reciprocal interaction between believing and feeling from a neurological perspective, Lect. Notes Artif. Int. {\bf 5819} (2009) 13.
\bibitem{9a} M. Hoogendoorn, S. W. Jaffry, and J. Treur, An Adaptive Agent Model Estimating Human Trust in Information Sources. In: R. Baeza-Yates, J. Lang, S. Mitra, S. Parsons, G. Pasi (eds.), Proceedings of the 9th IEEE/WIC/ACM International Conference on Intelligent Agent Technology, IAT'09. IEEE Computer Society Press, 2009, p. 458. 
\bibitem{10} K. Ku{\l}akowski,  Opinion polarization in the Receipt-Accept-Sample model, Physica A {\bf 388} (2009) 469. 
\bibitem{11p} G. Deffuant, D. Neau, F. Amblard and G. Weisbuch, Mixing beliefs among interacting agents, Adv. Compl. Sys. {\bf 3} (2000) 87.
\bibitem{11} K. Malarz, P. Gronek and K. Ku{\l}akowski,  Zaller--Deffuant model of mass opinion, submitted to JASSS ({\tt arXiv:0908.2519}). 
\bibitem{12} A. Sharpanskykh, Integrated modeling of cognitive agents in socio-technical systems, to be presented at KES AMSTA 2010.
\bibitem{13} K. Ku{\l}akowski and P. Gawro\'nski, To cooperate or to defect? Altruism and reputation, Physica A {\bf 388} (2009) 3581. 
\bibitem{14} M. Weber, Wirtschaft und Gesellschaft, J. C. B. Mohr (Paul Siebeck), T\"ubingen 1972.
\bibitem{15} Th. J. Scheff, Microsociology. Discourse, Emotion and Social Structure, The University of Chicago Press, Chicago 1990.
\bibitem{16} Report of the 7 July Review Committee of the London Assembly, Vol. 3: Views and information from individuals ({\tt http://legacy.london.gov.uk/assembly/reports/\-7july/vol3-individuals.pdf})
\bibitem{17} J. H. Turner and J. E. Stets, The Sociology of Emotions, Cambridge UP, Cambridge 2005.
\bibitem{18} S. Fortunato, Community detection in graphs, Phys. Rep. {\bf 486} (2010) 75.
\bibitem{19} M. E. J. Newman, Fast algorithm for detecting community structure in networks, Phys. Rev. E {\bf 69} (2004) 066133.
\bibitem{20} M. J. Krawczyk, Differential equations as a tool for community identification, Phys. Rev. E {\bf 77} (2008) 065701(R).
\bibitem{21} E. A. Leicht and M. E. J. Newman, Community structure in directed networks, Phys. Rev. Lett. {\bf 100} (2008) 118703.
\bibitem{22} J. H. Turner, Face-to-Face: Towards a Sociological Theory of Interpersonal Behavior, Stanford UP, Stanford, 2002.
\bibitem{23} K. Ku{\l}akowski, M. J. Krawczyk and P. Gawro\'nski, Hate--no choice. Agent simulations, in Psychology of Hate, ed. by C. T. Lockhardt, Nova Publ. 2010, in print.
\end{thebibliography}
\end{document}